\documentclass[preprint,5p,times,twocolumn]{elsarticle}

\usepackage{hyperref}

\usepackage{acronym}
\usepackage{bm} 

\usepackage{booktabs} 
\usepackage{amsmath}
\usepackage{amsfonts}

\usepackage{tabularx}
\usepackage{xcolor}

\journal{Engineering Applications of Artificial Intelligence}

\begin{document}

\acrodef{ML}{Machine Learning}
\acrodef{BSC}{Barcelona Supercomputing Center}
\acrodef{AIS}{Automatic Identification System}
\acrodef{IMO}{International Maritime Organization}
\acrodef{GPS}{Global Positioning System}
\acrodef{CRBM}{Conditional Restricted Boltzmann Machine}
\acrodef{MMSI}{Maritime Mobile Service Identity}
\acrodef{IMO}{ International Maritime Organization}
\acrodef{RBM}{Restricted Boltzmann Machine}
\acrodef{MSE}{Mean Squared Error}
\acrodef{SoG}{Speed over Ground}
\acrodef{CoG}{Course over Ground}
\acrodef{NRMSE}{Normalized Root Mean Squared Error}
\acrodef{ARIMA}{AutoRegresive Integrated Moving Average}
\acrodef{VHF}{Very High Frecuency}
\acrodef{GB-RBM}{Gaussian Bernoulli RBM}
\acrodef{MAE}{Mean Absolute Error}

\begin{frontmatter}

    \title{Improving Maritime Traffic Emission Estimations on Missing Data with CRBMs}

\author[BSC]{Alberto Gutierrez-Torre\corref{cor1}} 
\ead{alberto.gutierrez@bsc.es}
\author[BSC]{Josep Ll. Berral} 
\ead{josep.berral@bsc.es}
\author[BSC]{David Buchaca} 
\ead{david.buchaca@bsc.es}
\author[BSC]{Marc Guevara} 
\ead{marc.guevara@bsc.es}
\author[BSC]{Albert Soret} 
\ead{albert.soret@bsc.es}
\author[BSC,UPC]{David Carrera} 
\ead{david.carrera@bsc.es}

\address[BSC]{Barcelona Supercomputing Center, c/ Jordi Girona 1-3, 08034, Barcelona, Spain}
\address[UPC]{Computer Architecture department, Universitat Polit\`{e}cnica de Catalunya (UPC) - BarcelonaTech, c/ Jordi Girona 1-3, 08034, Barcelona, Spain}

\cortext[cor1]{Corresponding author.}

\begin{abstract}
Maritime traffic emissions are a major concern to governments as they heavily impact the Air Quality in coastal cities. Ships use the \ac{AIS} to continuously report position and speed among other features, and therefore this data is suitable to be used to estimate emissions, if it is combined with engine data. However, important ship features are often inaccurate or missing. State-of-the-art complex systems, like CALIOPE at the Barcelona Supercomputing Center, are used to model Air Quality. These systems can benefit from \ac{AIS} based emission models as they are very precise in positioning the pollution.  
Unfortunately, these models are sensitive to missing or corrupted data, and therefore they need data curation techniques to significantly improve the estimation accuracy. 
In this work, we propose a methodology for treating ship data using \acp{CRBM} plus machine learning methods to improve the quality of data passed to emission models that can also be applied to other GPS and time-series problems.
Results show that we can improve the default methods proposed to cover missing data. In our results, we observed that using our method the models boosted their accuracy to detect otherwise undetectable emissions. In particular, we used a real data-set of AIS data, provided by the Spanish Port Authority, to estimate that thanks to our method, the model was able to detect $45\%$ of additional emissions, representing 152 tonnes of pollutants per week in Barcelona and propose new features that may enhance emission modeling.

\end{abstract}


\end{frontmatter}

This version of document is the accepted manuscript, find the published article at {\color{blue}\url{https://doi.org/10.1016/j.engappai.2020.103793}}.

This manuscript version is made available under the CC-BY-NC-ND 4.0 license
{\color{blue}\url{http://creativecommons.org/licenses/by-nc-nd/4.0/}} \\
© 2020 Elsevier.

\section{Introduction}

Maritime traffic is considered an important contributor to primary atmospheric emissions in coastal areas~\cite{DiNatale2015166} and subsequently to European coastal air quality degradation~\cite{Viana201496}, especially in the North Sea and the Mediterranean basin. It has become a key component for European economy~\cite{ofxordeconomics15}, according to the European Community Shipowners Associations (ECSA) in 2015, being sea transportation more fuel-efficient than other modes of transport (e.g. trucks or trains). Nevertheless, according to recent reports by the \ac{IMO}, this form of transport will continue increasing in the future due to globalization and global-scale trade~\cite{smith2014}, increasing between $50\%$ and $250\%$ its contribution to the global Green-House Gas (actually $2.5\%$) by 2050. World governments, specially the European Union and the World Health Organization, are specially interested in advances on detection of emissions, for proper law enforcement towards the Air Quality Standards.

    Given this growing tendency, \emph{Smart Cities} need to be enabled to know how much pollution do the citizens suffer and act accordingly. In this case, ships can be seen as connected things like in the \emph{IoT} paradigm, informing about position and other characteristics with which exhaust emissions can be estimated. With all this information the cities will be able to evaluate pollution and propose more informed measures in order to have a healthier city.

Data from ship positioning and maneuvering is required to compute the emission components produced by maritime traffic and to understand levels of pollution from Environmental Research modeling techniques. This data is obtained from the \ac{AIS}, a \ac{GPS} based tracking system used for collision avoidance in maritime transport as a supplement to marine radars, providing for each vessel its unique identifiers, \ac{GPS} positioning and speed among other information. 

Well known and validated state of the art techniques to estimate emissions are using this information plus ship engine characteristics databases, like Jalkanen et al.~\cite{Jalkanen2009b, Jalkanen2012}. Doing so enables spotting sources and amounts of emitted pollutants. Such estimations result in large scale simulations and complex physics models, that feed from features like \emph{speed} and \emph{installed engine power}. Other techniques like the HERMESv3~\cite{guevara2019hermesv3} model in CALIOPE project~\cite{CALIOPE}, one of the trusted sources of Air Quality estimations by Spanish and Catalan Governments, performed at the Barcelona Supercomputing Center (BSC), use emissions reports, ship estimated routes and profiles by ship type for calculating ship emissions. While the HERMESv3 uses estimated routes, \ac{AIS}-based estimations can place the pollutants estimations accurately as it uses the actual ship placement, therefore improving the overall precision of the system.

However, \ac{AIS} data may be incomplete or faulty, e.g. information that could be used to enhance physical models like the ship operational mode (e.g. cruising) is incorrect, missing or poorly detailed too often, so are usually discarded when modeling emissions. This also may be the case of commercial databases that provide the required ship engine characteristics.

Further, dealing with \ac{AIS} data in populated regions is not trivial, as the average frequency of signal emission is of one message per each six seconds on average. Only in the coastal region of Barcelona represents 1.5 million registers per week. Processing this data in a periodical basis requires employing \emph{Big Data} techniques, understanding \emph{Big Data} as those situations where big volumes of input overwhelm our commonly used methodologies, making us to change them for techniques designed towards automation, scalability or approximation. Applying the complex physical models and simulations over such amount of data makes the problem more complex, requiring supercomputing infrastructures on a daily basis for periodical estimations, also predictions for public health and interest (CALIOPE computes 48-hour forecasts for all European continent).

To enhance such estimations, allowing better enabling features, we have available Data Mining and Machine Learning techniques, to refine, correct and fill missing data, allowing better accuracy on current air quality methods, also allowing experts on using once discarded features on physics models with higher confidence on the results. Data Mining provides consolidated techniques for analyzing such data, extracting relevant values, frequent and rare patterns, and also model behaviors. Most of those wanted patterns are not trivial or present at simple sight, even they can be found across huge amounts of data, unable to be handled exclusively by human experts. Considering the \ac{AIS} obtained profiles for each ship of any size and characteristics as multi-dimensional time-series representing their behavior, we can discover new latent features that can enhance the \ac{AIS} data-set towards modeling emissions. There are several approaches for mining patterns on time series, e.g. stream mining methods for time-changing data~\cite{Bifet07learningfrom}, series-aware neural networks as \acp{CRBM}~\cite{Taylor2009} or \emph{Recurrent NNs}, even Hidden Markov Models for time-series modeling~\cite{Ghahramani:2001:IHM:505741.505743}.

This work provides a methodology to enhance the obtained \ac{AIS} data-sets by cleaning, treating and expanding some of its features using domain knowledge, to produce better emission models and correct emission inaccuracies. The proposed methodology focuses on using \acp{CRBM} to boost clustering and prediction algorithms, to improve the quality of features like ship main engine power, navigation status and ship category, from each ships navigation traces. The decision of using \acp{CRBM} is based on their capacity to deal with \ac{AIS} as multi-dimensional time-series~\cite{conf/uai/MnihLH11}, also encouraged by the methodology proposed by Buchaca et al.~\cite{buchaca} used for detecting phase behavior patterns on time-series. The \acp{CRBM} are used to extract and cluster temporal patterns, also to expand features from the time series, allowing non-time-aware predictors better accuracy. Our methodology combines \acp{CRBM} with clustering techniques (i.e. $k$-means) and prediction techniques (e.g. Random Forests, Gradient Boosting, Lasso) towards predicting and characterizing the engine installed power, the navigation status and ship types from their traces, for vessels do not have any of those attributes or that provide them incorrectly.

To summarize, the contributions are the following:
\begin{enumerate}[ 1{)} ]
    \item Generate feature representations of local behavior of ship movements using \acp{CRBM}. This new feature representation is a key building block for \ref{contrib2}) and \ref{contrib3}).
    \item \label{contrib2} Ship type and main engine installed power missing values correction for better emission estimations using the previously generated features plus machine learning algorithms.
    \item \label{contrib3} Provide an initial step to correct and improve Navigational Status AIS feature, using the generated features plus a clustering technique. 
\end{enumerate}

The current approach has been tested using real \ac{AIS} data provided by the Spanish Ports Authority (\emph{Puertos del Estado}), complemented by ship and engine characteristics coming from the Ship database provided by IHS-Fairplay. We validate and test our approach to enrich and complete data by comparing it against the current state of the art approaches with a methodology based on Jalkanen et al.~\cite{Jalkanen2009b, Jalkanen2012} in scenarios of missing data, in a supervised manner to get emissions computed with real and complete data. After the aggregation of the emissions, we show that when enhancing the data with our method we are able to detect a $45\%$ of the previously undetected emissions (152.95 tonnes out of 343.47) when applying the proposed standard procedure. We also show how other faulty features like Navigation Status and Ship Type can be corrected or improved.

This article is structured as follows:
Section \ref{sec:related} recaps the related work and current state-of-art on \ac{AIS} applications, treatment on time series, \acp{CRBM}, and also related studies on maritime pollution.
Section \ref{sec:background} introduces some background on the \ac{AIS} mechanisms and usages, also introduces the fundamentals of \acp{CRBM}.
Section \ref{sec:preparation} describes the data-set, its features and details.
Section \ref{sec:methodology} explains our methodology and machine learning pipelines.
Section \ref{sec:experiments} details the experimentation and validation of our methodology.
Finally, section \ref{sec:conclusions} summarizes this work, and presents the future work.


\section{Related Work}
\label{sec:related}

AIS-assisted emission estimations can be effectively used to assist policy design and corrective measures of a specific shipping sector (e.g. cruises and ferries)~\cite{RePEc:eee:transa:v:78:y:2015:i:c:p:347-360} and to improve the efficiency of ships~\cite{Gao_Makino_Furusho_2016}. Also, works by Jalkanen et al.~\cite{Jalkanen2009b,AMBIO-Jalkanen2014} show that AIS data can been used for the estimation of high spatial and temporal resolution maritime emissions. Compared to traditional emission estimation methodologies, the use of AIS data provides information of instantaneous speed, position and navigation status of vessels and subsequently allows for more accurate estimations of vessels' activities and the improved reliability of emissions and fuel consumption estimations~\cite{buhaug2009}. Navigational status is included on \ac{AIS} data and with this attribute the current engine usage can be estimated along with other attributes like speed, however in some cases it is incorrectly set as this attribute is manually set.

One of the main issues when using \ac{AIS}, highlighted by Miola et al.~\cite{miola2010} are data gaps and anomalies. In certain occasions not all of the data fields are fully or correctly populated (e.g. navigation status is incorrectly reported or just missing, speed calculated from real AIS information reaches unrealistic values). This affects the suitability of raw \ac{AIS} data for estimating air emissions around ports. Furthermore, other ship characteristics needed for the emission modeling (e.g. installed engine power) may also be missing. In this work, we propose the use of machine learning as a solution to correct and refine that data, specifically \acp{CRBM}.

Pattern mining for GPS traces is a common practice in very different fields, looking for specific patterns in movement and behavior. Works like Qiu et al.~\cite{7194272} describe a methodology for mining patterns through Hidden Markov Models, producing semantic information to feed frequent pattern mining methods. Such work is also based on discovery of frequent episodes in time series~\cite{Mannila:1997:DFE:593416.593449}, with the goal of discovering patterns series of events. Use cases for such techniques are social mining and recommendation~\cite{Zheng:2010:CLA:1772690.1772795}, animal movement patterns~\cite{Li:2011:MMM:1989734.1989741}, or elder care~\cite{6485238}. Here we proceeded to find common patterns using \acp{CRBM} as a base for time windows, feeding them from GPS and other input sources, for discovering discriminating behaviors on a geographical space.

The \acp{CRBM}, as probabilistic models derived from \acp{RBM}~\cite{conf/ciarp/FischerI12, series/lncs/Hinton12}, are used in a wide range of problems like classification, collaborative filtering or modeling of motion capture, developed by the team of professor Geoffrey E. Hinton at the University of Toronto~\cite{conf/uai/MnihLH11, Salakhutdinov:2007:RBM:1273496.1273596, Taylor:2009:FCR:1553374.1553505, Taylor06modelinghuman}. Such models are usually applied for problems where time becomes a condition on data, i.e. time-series. Other works like X.Li et al.~\cite{DBLP:conf/aistats/2015} and Lee et al.~\cite{Lee:2016:AMC:2910739.2910812} use the models for multi-label learning and classification. Based on their experiences and techniques, we are taking advantage on \acp{CRBM} time-series learning capabilities.

\section{Background}
\label{sec:background}

\subsection{AIS and the CALIOPE Project}

\begin{figure}[h!]
	\centering
    \includegraphics[width=1\linewidth]{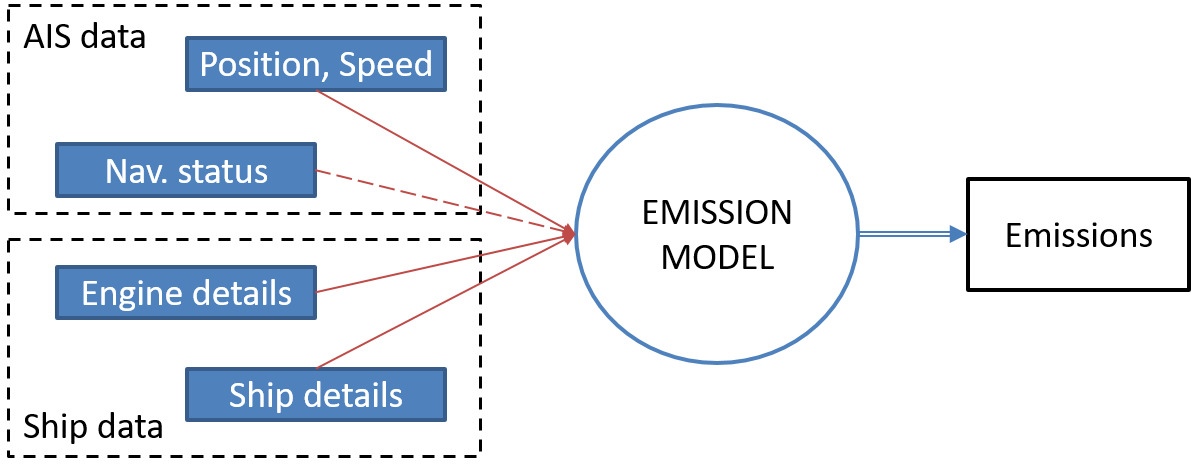}
	\vspace{-3mm}
	\caption{Estimation of Emissions from Ship Traces}
	\label{fig:estimation_diagram}
\end{figure}

Ship traces can be obtained from the \ac{AIS}, the \ac{GPS} based tracking system used for collision avoidance in maritime transport as a supplement to marine radars. This used to track and monitor vessel movements from base stations located along the coast, and transmitted through standardized \ac{VHF} transceivers. \ac{AIS} provides for each vessel its unique \ac{IMO} identifier and \ac{MMSI} number, \ac{GPS} positioning, also course and speed among other information. This system is mandatory, according to \ac{IMO}'s \textit{Convention for Safety of Life at Sea}, for all ships with gross tonnage greater than 300 tons, and all passenger ships~\cite{imoais2017}. For this reason, this data has become relevant on air quality studies, as summarizes Figure~\ref{fig:estimation_diagram}.

The CALIOPE air quality forecasting system~\cite{CALIOPE} is a state-of-the-art modeling framework that integrates a meteorological model, an emission model, a Saharan dust model and a chemical transport model to simulate air quality concentration with a high spatial (up to 1km$^2$) and temporal (1 hour) resolution. CALIOPE is currently used by Spanish air quality managers, like the Generalitat de Catalunya, for environmental policy making. Air quality results are continuously evaluated with a near real time system based on measurements from the Spanish air quality network, and the performance of the system has been previously tested in different evaluation and air quality management studies~\cite{Soret201451}. The HERMESv3 model is the emission core of the CALIOPE system and has been fully developed by the Earth Science department of the \ac{BSC}~\cite{12857432}. Due to high impact of maritime traffic on ambient pollutant levels at the urban area of Barcelona~\cite{27474834} one of the current objectives of the group is to improve the emission estimation of this activity using an AIS-based methodology. A collaboration has been set up with \ac{BSC} Earth Sciences Department in order to tackle this problem.

Our approach is a methodology based on the STEAM model, proposed by Jalkanen et al.~\cite{Jalkanen2009b,Jalkanen2012}, is an \ac{AIS} based emission estimation model that uses the traces of the ships and their characteristics to provide information about the pollution with \ac{GPS} positioning precision. Emissions are calculated using the current power consumption of the ship at a given time and the emission factor for that ship regarding a pollutant. Conceptually, the formula is the following (units in brackets):
$$E_{s,p,x,t}[g/h] = P_{s,x,t}[kW] \cdot EF_{s,p}[\frac{g}{kWh}]$$

Being $P$ the current power consumption of the ship and $EF$ the emission factor, $s$ ship characteristics (mainly engine), $p$ the pollutant to estimate, $x$ position of the ship and $t$ the current time. Therefore, the emission of a given pollutant for a ship that is in the position $x$ and time $t$ is conditioned by the ship characteristics for calculating the actual power that this engine is using and the ship characteristics plus the pollutant constants for the installed engine.

Generally, in ships there are two kinds of engines installed: main engine and auxiliary engine. The first is mainly in charge of the movement of the vessel and the latter is mainly in charge of the on-board electrical power devices but may be used for other tasks. 

The power used at a given time $t$ on the main engine is estimated using the current vessel speed, the design speed and the installed power. In particular, the formula to calculate the transient main engine power is the following:

$$ P_{transient} = \frac{V^{3}_{transient}}{(V_{design} + V_{safety})^3} * \varepsilon_{p} * P_{installed} $$

Being $V_{transient}$ the current speed provided by \ac{AIS}, $V_{design}$ the maximum speed that the ship can reach by design, $V_{safety}$ a safety offset as ships may report speeds slightly greater than $V_{design}$, $\varepsilon_{p}$ engine load at Maximum Continuous Rating and $P_{installed}$ the actual power installed in kilowatts.

Following the instructions from STEAM methodology, $V_{safety}$ is fixed to 0.5 knots (2,57 m/s) and $\varepsilon_{p}$ is set to 0.8. 

When installed power and design speed are missing, Jalkanen et al.~\cite{Jalkanen2009b} propose to use the average of those characteristics for the given ship type, usually available in \ac{AIS} data. However, the installed power has high variance giving an estimator of the installed power with low accuracy. As design speed can be obtained directly from installed power, we focus on the main engine power.

In the case of auxiliary engine, the methodology defines 3 stages: hoteling (moored), maneuvering and cruising. These three operational modes have a power consumption associated, different by type, as there is no other information available for the power estimation of this type of engine. The operational modes are assigned to the trace using the current speed of the ship.
Regarding that for each stage we have a different value for a given ship type, it is very important to know the actual ship type. In some cases this type is incorrect or missing in \ac{AIS} data.

Also, \ac{AIS} data provides the navigational status of the ship. However, this attribute is not reliable as it is manually set by the crew and it can have delays or be incorrectly set, e.g. in fishing ships the status is always \emph{engaged in fishing} even if they are hoteling. This attribute can potentially give more information about the current usage of the engine than the simpler division done using the speed of the ship, however it needs to be fixed in some cases.

Given these shortcomings we propose a methodology in order to improve the quality of each attribute, being the main engine power and ship type the ones that can be apply with the current model and navigational status cleaning and expansion as an enabler for a more precise future emission model step.

\subsection{Conditional Restricted Boltzmann Machines}
 
\subsubsection{Restricted Boltzmann Machines}

A \ac{RBM}, or more concretely \ac{GB-RBM} is a key building blog of the \ac{CRBM}. A \ac{GB-RBM} is an undirected graphical model with binary hidden units and visible Gaussian units that models the joint log probability of a pair of visible and hidden units $(\bm{v},\bm{h})$ as
\begin{equation}
\log P(\bm{v},\bm{h}) = \sum_{i=1}^{n_v}  \frac{(v_i -c_i)^2}{ 2 \sigma_i^2} - \sum_{j=1}^{n_h} b_j h_j - \sum_{i=1}^{n_v} \sum_{j=1}^{n_h} \frac{v_i}{\sigma_i} h_j w_{ij} + C 
\end{equation}
where $\sigma_i$ is the standard deviation of the Gaussian for visible unit $i$, $c$ is the bias of the visible units , $b$ is the bias of the hidden units, $w_{ij}$ is the weight connecting visible unit $i$ to hidden unit $j$ and C is a constant. Notice that $n_v$ and $n_h$ refer to the dimension of  $\bm{v}$ and $\bm{h}$ respectively. In practice we normalize the data to have zero mean and unit variance. Moreover, $\sigma_i$ is fixed to 1 because it empirically works well as shown in the work of Taylor et al. \cite{Taylor06modelinghuman}.


\subsubsection{Conditional Restricted Boltzmann Machines}

The \ac{CRBM} is a GB-RBM that models static frames of the time series  modified with  some extra connections used to model temporal dependencies. The CRBM keeps track of the previous $n$ visible vectors in a $n \times n_v$ matrix which we call the history of the CRBM.  The learned parameters of the CRBM are three matrices $\bf{W}, \bf{A}, \bf{D}$, as well as a two vectors of biases $\bf{c}$ and $\bf{b}$ for the visible and  hidden units respectively.  ${\bf W} \in   \mathbb{R}^{n_v \times n_h}$ models the connections between visible and hidden units. ${\bf A}  \in \mathbb{R}^{ (n_v \cdot n) \times n_v}$  is the mapping from the history to the visible units. ${\bf D}  \in \mathbb{R}^{(n_v  \cdot  n) \times n_h}$ is the mapping from the history to the hidden units. 
Inference in the CRBM is performed using the contrastive divergence method. More information can be found in Taylor et al.~\cite{Taylor06modelinghuman}.



\begin{figure}[ht!]
	\centering
    \includegraphics[width=0.8\linewidth]{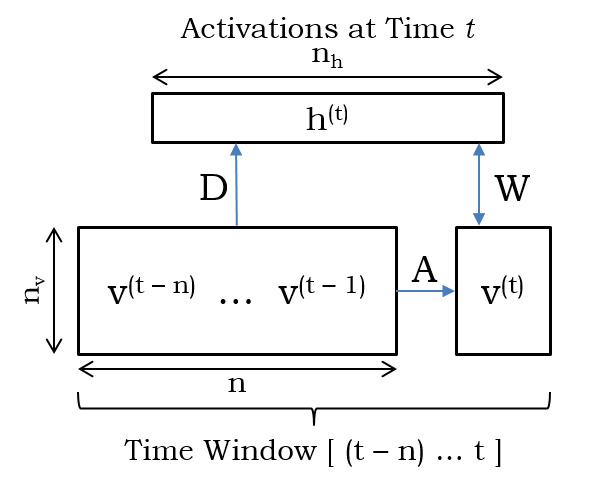}
	\caption{CRBM with window size $n$ and $n_h$ hidden units}
	\label{fig:crbm_diagram}

\end{figure}

    Figure~\ref{fig:crbm_diagram} shows a graphical representation of a CRBM. In the case of this paper, we have interest on using the activations produced by the hidden units when feed with a sample of ship traffic trace plus the $n$ steps window. Then we use this vector of activations as a time-aware code that represents a sample with a time window for algorithms that are not thought to handle time-series, enabling them to manage temporal dependencies. The actual input data for the \ac{CRBM} is defined in Section~\ref{sec:preparation}.

Notice that even though \ac{CRBM} suffer when doing long-term predictions, this fact does not affect the taken approach as it is being used as an encoder and not a predictor.

\section{Data Preparation}
\label{sec:preparation}

\subsection{Data-set Properties}

The current data-set has been provided by the Spanish Ports Authority (\emph{Puertos del Estado}), from their vessel monitoring database collecting the \ac{AIS} signals from all registered ships navigating national waters. Such database collects the information periodically sent from all registered vessels, and can be used by local port authorities. The data-set used for current experiments is a slice of data concerning the coastal area of Barcelona, including a week of maritime traffic. It is composed by more than 1.5 million entries and indicating 19 features. The relevant variables of the dataset for this study will be introduced later on.

\emph{Puertos del Estado} has deployed a network of AIS base stations through the whole Spanish coast, with the dual objective of obtaining maritime traffic information (especially at the port area) and applying the AIS capabilities to navigation aid\footnote{http://redais2.puertos.es}. Each AIS base station is responsible for receiving the AIS data within its coverage area and sending it to the central hub for processing, storage and subsequent distribution to other AIS networks or interested users.

Each vessel is identified by 1) name of the ship, 2) the \ac{IMO} number, given by the \ac{IMO}, 3) and the \ac{MMSI} number. There are two \ac{AIS} device classes (A and B) differing in transmission power and capabilities, being Class B smaller and short ranged than Class A. Ships transmitting with a Class B device are not required to have an \ac{IMO} number, then having \emph{Not Available} values (NAs) in our data. \ac{MMSI} is used as identifier if it is not explicitly defined in each case. Moreover, \ac{AIS} devices are periodically transmitting \emph{static} attributes, properties of the ship that do not change on time, e.g. length, beam or draught, so authorities and other ships are able to know the size of the vessel. These last attributes are not considered for this study as they are unreliable for the current task.

On the other hand, from the dynamic data provided by \ac{AIS}, the following subset is used in this study:
\begin{itemize}
\item Time-stamp of the transmission
\item \ac{GPS} Coordinates in latitude-longitude
\item \ac{SoG}: Speed of the boat, measured as effective over ground, by taking into account the tidal drifting or speeding up/down the ship, measured in knots.
\item Navigation Status (navstatus): A standardized identification of the current status of the ship. This feature is manually set by the crew. This denotes the susceptibility of such feature to errors and missing values.
\item Type of ship and Cargo (typeofshipandcargo): A combination of two integer values, encoding the type of ship and materials that it is currently transporting.
\end{itemize}
Additionally, the \ac{AIS} provides information like the ship rotation (\ac{CoG}), the rotation speed and compass heading. These features have proved to be unstable to perform accurate predictions. The information from each single vessel is collected in their navigation trace along time. Table~\ref{tab:raw_data} shows a sample from our data-set.

\begin{table*}[ht!]
\centering
\resizebox{\textwidth}{!}{%
\begin{tabular}{rrrrrrrrrrrrrrrrr}
  \hline
  ID& size\_\{a, b, c, d\} & length & beam &draught & sog & cog & rot & heading & navstatus  & type & lat & lon & timestamp\\       
  \hline                                                   
  1 &  62 , 126 ,  13 ,  15 & 188 &  28 & 7 & 5.50 & 317 & 127 & 326 &   0 &  70 & 40.91 & 2.47 & 2014-04-13 23:59:32 \\ 
  2 &  17 ,  19 ,   7 ,   1 &  36 &   8 & 3 & 0.00 & 170 &   0 &  47 &   8 &  37 & 41.53 & 2.44 & 2014-04-13 23:59:31 \\ 
  3 &   4 ,  16 ,   4 ,   2 &  20 &   6 & 4 & 10.00 & 220 & -128 & 511 &   7 &  30 & 41.30 & 2.19 & 2014-04-13 23:59:33 \\ 
  \hline
\end{tabular}
}
\caption{Sampled data from the data-set. Identifiers are surrogates from the real identifiers}
\label{tab:raw_data}
\vspace{-5mm}
\end{table*}

\subsection{Cleaning and Normalizing Data}
\label{sec:clean}
Working with time-series implies having data regularized in time, as many techniques interpret samples as steady and regular, more than sparse, occasional or even redundant. When using \acp{CRBM} with time as conditioner, each position in the \emph{delay} (the window of data history) is supposed to be given a set of weights towards the hidden layer, then data values slide through the window facing new weights based uniquely on their position in history. This way, each position in the history window discretizes time in equal segments, so sparse data needs to be densified, and missing data must be interpolated or predicted.
%
%
In order to do this, linear interpolation is applied to adjust data points to a regular time scale, as performed in the previously mentioned studies by Jalkanen. Even though more advanced interpolation algorithms can be used, we have chosen to follow the linear interpolation procedure described by Jalkanen so that the results are standard.


\subsubsection{Cleaning and interpolation}

First of all, rows with incorrect time-stamps are removed if there is no possibility of repairing them. 
Names, if missing, are obtained from third party ship databases, freely providing such information, such as VesselFinder~\cite{vesselfinder}.
%
%
Next step is to retrieve each ship time series and then processed it for time regularization and interpolation when required. Data goes through a two step procedure: 1) Produce the time-steps for the desired time granularity, e.g. a sequence of steps of 5 seconds in between; and 2) Using the available data, linearly interpolate the steps generated in the previous operation if the time difference between samples is less than 72 hours\cite{Jalkanen2009b}.


In order to avoid bias or over-fitting on locality when searching for patterns, a new feature is added indicating the relative movement, by obtaining the difference in Latitude/Longitude between each consecutive points. This way we register the movements between registered observations instead of absolute values, having a movement feature free of geographical information. Also, the same procedure can be performed over rotation features, having as result relative rotation movements. However, rotation attributes from \ac{AIS} are not always available, hence here we created a rotation variable calculated from the \ac{GPS} traces as they are more reliable.

Another generated feature is the zone location of vessels. Following the information provided from CEPESCA~\cite{Cepesca}, we consider that sea is divided in three zones: coast, fishing area and high sea. These zones respond where bathymetry is below 50 meters, not suitable for fishing and close to coast, between 50 and 1000 meters, where fishing vessels labor, and beyond 1000 meters as high sea.

After pre-processing we have a time series for each ship, with regularized time-steps between observation, and new derived features indicating relative positions and movement, allowing us to compare ships for their positioning and maneuvering, independent of the origin port or coastal point, even from length of some pattern repetitions. 

The final features, from now on called \emph{original features}, used for training the \ac{CRBM} and for comparison in the experiments are the following:
\begin{itemize}
    \item Ship ID: IMO and MMSI number
    \item Ship type
    \item Relative GPS rotation 
    \item Speed over Ground
    \item Bathymetry as a 3 value categorical variable: coast, sea and high sea
\end{itemize}

In Section \ref{sec:usecase4} a new dataset will be introduced and used for that particular task along with the previously defined data.

\section{Learning Models}
\label{sec:methodology}

Our methodology proposes a set of learning pipelines towards improvement of emission estimation. First, we attempt to reconstruct the missing data on ship type and engine power, required for emission estimations. The pipeline for this estimation consists on using the \acp{CRBM} to produce a new set of latent features, to be ingested by classification (ship type) and regression (engine power) algorithms. Second, we have the pattern mining ensemble, using the \acp{CRBM} along clustering algorithms (i.e. k-means), to find the NavStatus patterns and behaviors on ship traces, intended for future emission models where NavStatus and latent sub-type ship data can be used.




\subsection{Data Pipelines}

\begin{figure}[h!t]
	\centering
    \includegraphics[width=\linewidth]{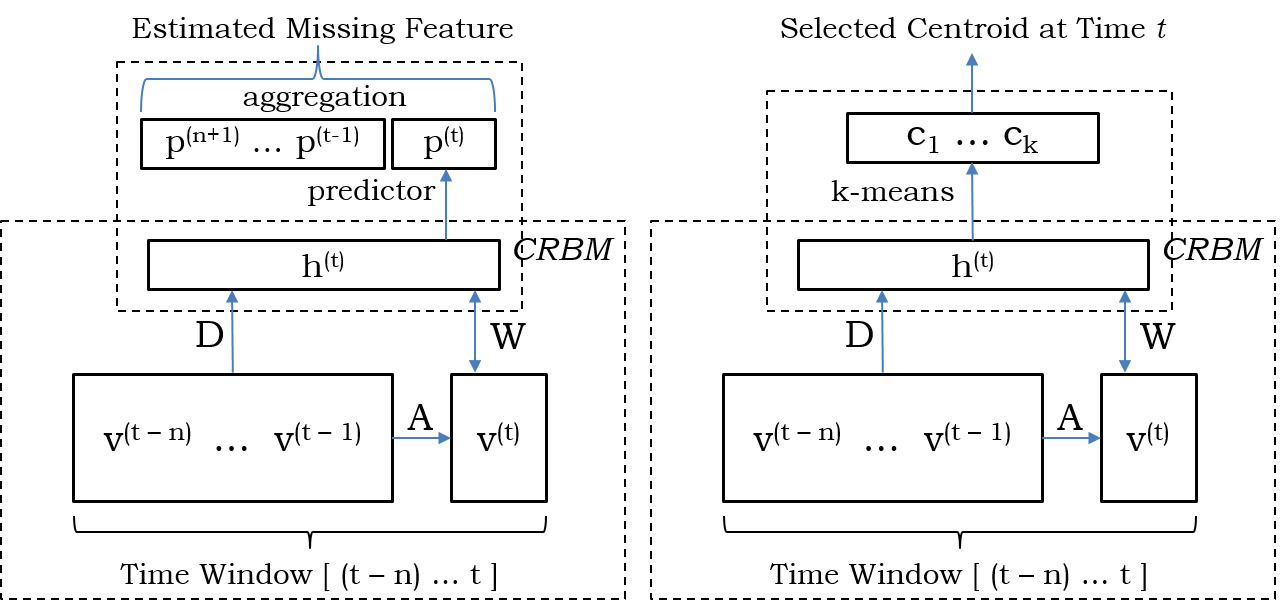}
	\vspace{-6mm}
	\caption{Pipeline for CRBMs, Prediction and Clustering}
	\label{fig:crbmpipeline}
	\vspace{-4mm}
\end{figure}

The \acp{CRBM} require as input a time-series window, fed by the traces for each ship. The data-set is composed by time-windows of size $n+1$, considering at each time $t$ an Input $v^{(t)}$ of dimensions $n_v \times 1$, and a history record $v^{(t-n)} \ldots v^{(t-1)}$ of dimensions $n_v \times n$. This data-set is produced by sliding the time-window from time $n$ to $T$, where $T$ is the total number of recorded steps. Notice that this implies to \textit{burn} the first $n$ records in order to create a history record for Input $v^{(n+1)}$. The outputs of the \acp{CRBM} are a vector of size $n_h$, where $h$ is the total size of generated latent features. This \ac{CRBM} is connected to a prediction algorithm that will feed from those features, from now on called \textit{Activations}, and compared with the real output variables (ship type and engine power) in a supervised learning fashion. For the scenario of correcting the NavStatus and characterize ships for further usage, we connect the \acp{CRBM} to a clustering method \emph{k-means}, fed by the \ac{CRBM} activation vector. The idea of using \acp{CRBM} rely on the fact that those mentioned classical learning algorithms are fed with implicitly time-aware data. Figure~\ref{fig:crbmpipeline} shows the schema for both of the pipelines.

\subsection{Training the Pipeline}

The \ac{CRBM} is trained with sample series of data, structured as explained previously. Providing the history record matrix to a \ac{CRBM} as the \textit{Conditional element}, provides the notion of time. This allows training it through data batches without forcing any particular order between batches, as the notion of order is present within each batch. Best practices in modeling and prediction require to split training data with validation and testing data, to prevent the auto-verification of the model, so for this reason we performed this training process with a subset of the available time-series. Also the splits have been performed using the \emph{ids} from the traces (each \emph{id} identifies a single time-series), therefore none of the splits shares a single time step from the other split. Each instance passing through the \ac{CRBM} is encoded into an activation vector of size $n_h$. This way, the ship tracking information and history are codified by a $n_h$-length vector, knowing that as far as a CRBM reconstruction misses the original data by little, such vector contains a compressed or expanded version (depending on the values of $h$ and $n$) of the current and historical status of such ship.

After training the \ac{CRBM}, using the activation data-set we train the prediction and clustering algorithms. The principal hypothesis is that ships with similar properties will produce similar activations over time. For the prediction scenarios, we are using well known algorithms like Random Forests, Gradient Boosting, Multi-layer perceptron networks, Logistic Regression and Lasso. We chose those models as the ones performing better or the best models for baseline comparison, discarding those performing worse on accuracy and slower on training stages. As we have a set of activations, from time $n+1$ to $T$, for each output value $P$, each ship traces pass through the pipeline creating a vector of predictions $\hat{p}^{(n+1)} \ldots \hat{p}^{(T)}$, then aggregated and compared against $P$. For classification we used the fashion (top vote), while for regression we used average and median, as explained on the experiments. For the unsupervised scenarios, we focused directly on k-means algorithm for its simplicity of use and interpretation. We experimented with these models and their hyper-parameters using cross-validation and grid search, on a 6 Intel Xeon 40-core and 128GB RAM cluster.

\section{Experiments}
\label{sec:experiments}

The experiments focus on the different pipelines used towards the improvement on emission estimations, as follows:

\begin{enumerate}
	\item Discriminate ship types: predict the type of a ship given its behavior at every time step.
	\item Improving main engine emission modeling: predict installed main engine power when missing values for emission estimations.
	\item Navigational status pattern mining: determining the status of vessel directly from reliable GPS coordinates, potentially correcting badly input \emph{NavStatus} values and finding uncovered behaviors.
\end{enumerate}

After several experiments with feature selection and refinement of aggregated features, we selected as input features the \emph{bathymetry}, the \emph{\ac{SoG}}, and the \emph{GPS-rotation} (the rotation calculated from the GPS positioning, as the feature \emph{rot} is frequently missing or with incorrect values), as mentioned in Subsection~\ref{sec:clean}. Bathymetry is indicative of the geographical zone where vessels are navigating, if coastal zones, fishing zones, and open sea. Speed and rotation provide the vector of the vessel movements.

Following the proper methodology for training models, we have separated the data-set into training and test, by a random split $0.66-0.34$ of the ship series. To measure the \ac{CRBM} errors (minimal error at data reconstruction), we used the testing series and performed a \emph{simulation}: passing the whole series through the \ac{CRBM} for activation and reconstruction (i.e. the process of generating the input features from the activations), then computing the \ac{MSE} of the inputs and reconstructions.

During the experiments we attempted different \ac{CRBM} hyper-parameters, with a wide range of hidden units in the hidden layer, and different \emph{delay} or history window length. For most of the experiments, we concluded that 10 hidden units with 20 observations of history (1 per minute), provided us the best reconstruction results and differentiating clusters. For the experiment in Subsection~\ref{sec:usecase4} we concluded that expanding features (from 60 visible units to 70 hidden units) produced better prediction results.

\subsection{Ship Type Prediction}
\label{sec:usecase1}

In the emission estimation process it is required to know the type of the ship as the emission model use it to determine how much power is the auxiliary engine producing given the navigation status. For instance, cruisers are always assumed to constantly use 4000kW of auxiliary engine power. On the other hand, other ship types may be estimated to use 750kW during cruise, 1250kW during port maneuvers and 1000kW while hoteling. Therefore it is very important to know which type the ship is to correctly asses the auxiliary engine consumption.

The following experiment consists on using the \ac{CRBM} classification pipeline for classifying ship types on a supervised learning scenario. A train/test split of $70\%-30\%$ is used along with cross-validation for hyper-parameter search. The target class attribute \emph{type} is retrieved from the \emph{typeofshipandcargo} \ac{AIS} variable, that encodes it on its first digit. 
We are dismissing the type of cargo at this moment, and focusing on the type of ship. However this makes some classes to be more difficult to predict, e.g. class 3 ``Special Category'' has a range of different ships from fishers to tugs, therefore this is a hard classification problem due to the heterogeneity of ships within one class.


As classifying algorithms we used Logistic Regression, also Multi-Layer Perceptron network with different values of neurons at the hidden layer, from 100 to 500, obtaining the best results with 500 hidden units, a rectification layer and the \emph{Adam} optimization method, with 0.9 momentum. Models have been trained and evaluated with three different sets of features as its input: 1) the original features (\emph{bathymetry}, \emph{SoG}, \emph{GPS-rotation}); 2) the original features appended by a time window of 20; and 3) the 10 hidden unit activations of the CRBM.

\begin{figure}[h!]
	\centering
	\noindent\makebox[\linewidth]{%
        \includegraphics[width=1\linewidth]{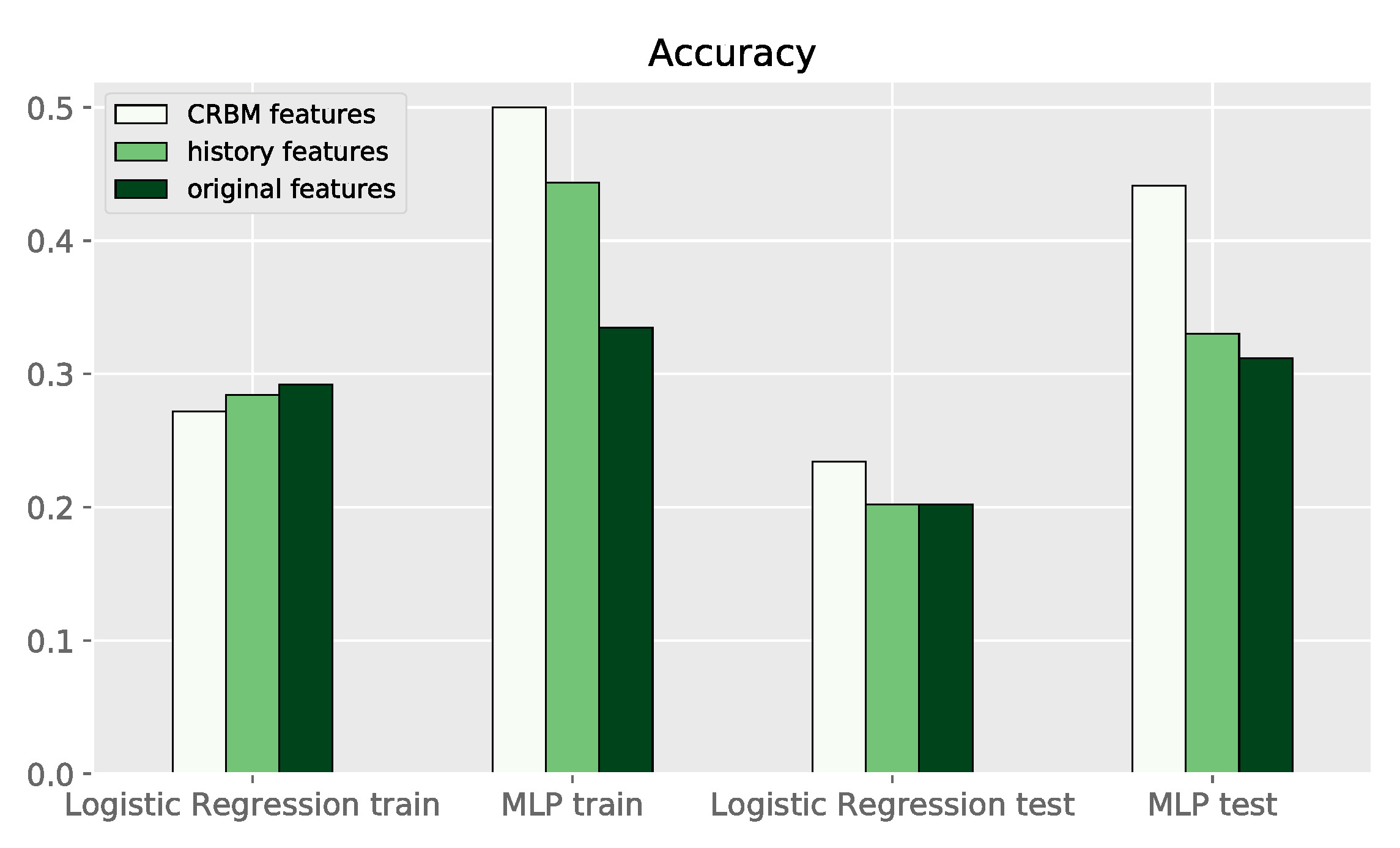}
	}
	\vspace{-3mm}	
	\caption{Accuracy of the ship classification pipeline depending on the features. }
	\label{fig:train_test_acc_ship_classification}
\end{figure}

\begin{figure}[h!]
	\centering
    \includegraphics[width=\linewidth]{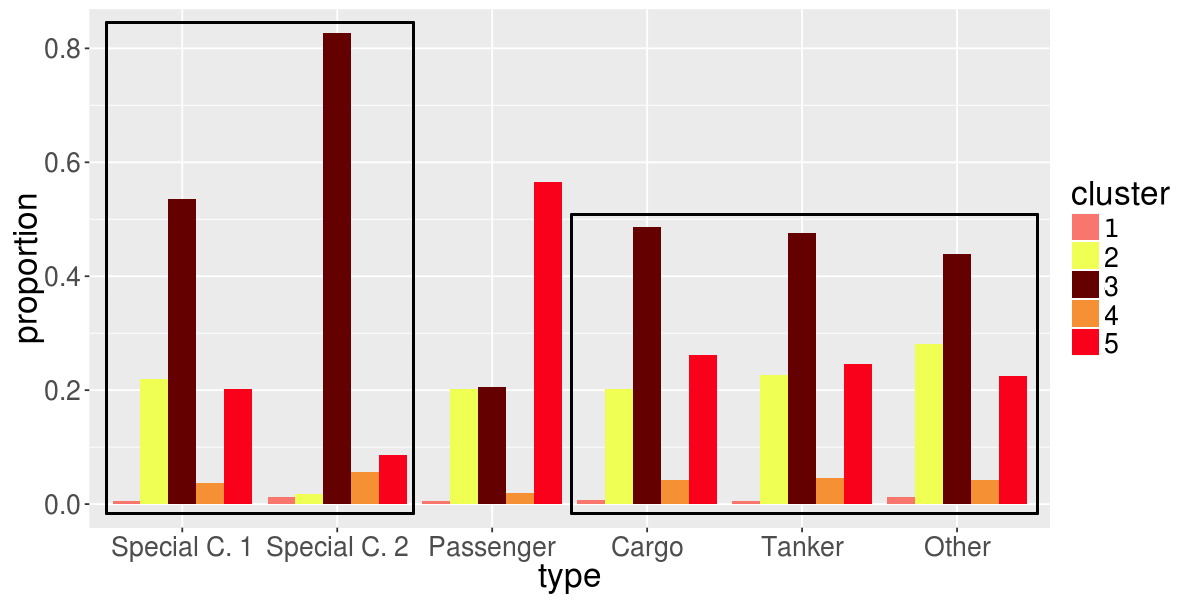}
	\vspace{-2mm}
	\caption{Proportion of patterns by ship type. It can be seen that some ship types have similar pattern frequency, but others differ drastically. For example, both special categories 1 and 2 (class 3 and 5 respectively) are quite different as 3 is more fishing oriented than the service oriented class 5. On the other hand, carrier ships and others are similar.}
	\label{fig:propTYpe}
\end{figure}

Figure~\ref{fig:train_test_acc_ship_classification} shows the accuracy result  on the train and test sets. The \ac{CRBM} features helped both models to
identify ship types and produced more balanced models, e.g. in the case of logistic regression the other features get better results in
training, but in test the activations are better and closer to the training
error. This can also be seen in Figure~\ref{fig:propTYpe}, in which it can be seen that the proportion of \emph{k-means} clustering result, i.e. the proportion of the found patterns inside one class, is different in some ship type, therefore it seems that \ac{CRBM} enhances data separability for this case.

\subsection{Improving main engine emission estimations on the presence of missing data}
\label{sec:usecase4}

In order to estimate the emissions that a ship produces some characteristics of the engine are required, as shown in the work of Jalkanen et al.\cite{Jalkanen2009b, Jalkanen2012} with their STEAM methodology. In particular, one of the most important attributes is the main engine power. This data is available in commercial databases provided by companies, e.g. IHS Fairplay~\cite{IHS_FAIRPLAY}, however the data may be missing for some ships. 
%
%
In this case the aim is to provide correct attributes to an estimation model results that have been already validated when all the information is correctly given. When there are missing attributes, assumptions need to be made. In case that there is no data about the ship aside from the data given by \ac{AIS}, the suggested approach in STEAM methodologies for main engine power is to use the average by ship type. It is a simple and effective solution but not the best as the variance in installed power by ship type is high. We want to use the ship trace along the ship type, data provided by \ac{AIS} to estimate the ship characteristics needed for pollution estimation, in this case the already mentioned main engine power. 


The experiment has the same setup as the use case of Subsection~\ref{sec:usecase1} in terms of validation framework. In this case, the \ac{IMO} number is used as identifier, as the IHS data-set only contains ships with ``class A'' transceivers and does not provide an \ac{MMSI} field, hence we have to discard all the ships that do not provide an \ac{IMO} number for this experiment. We predict the main engine and we validate it using a separated test set. We have trained a new CRBM model specifically for this use case. The best results were obtained with 70 hidden units and with ensemble vote function \emph{median}, which is more stable than the mean in the experiments using the \ac{CRBM} activations.



For this experiment we tested different algorithms, e.g. Random Forests, Gradient Boosting and Lasso Regressing, the average of values per type as proposed by Jalkanen et al.~\cite{Jalkanen2009b}, and the global average. The model that produced the best result was Random Forest with 200 estimators and no limit in number of features and depth, as can be observed in Table~\ref{tab:MAEpowerME}. 

\begin{table}
\centering
\begin{tabular}{lrrrr}
\toprule
&  \multicolumn{2}{c}{Median} &   \multicolumn{2}{c}{Mean} \\
    Model &  \multicolumn{1}{c}{Train} & \multicolumn{1}{c}{Test} & \multicolumn{1}{c}{Train} & \multicolumn{1}{c}{Test} \\
\midrule
Global avg  &  8815.28 &  13279.71 &  8815.28 &  13279.71 \\
Type avg    &  7904.08 &  11870.83 &  7904.08 &  11870.83 \\
Lasso act.  &  8562.75 &  12744.20 &  8562.75 &  12744.20 \\
Lasso hist. &  8430.93 &  12933.68 &  8237.30 &  12616.83 \\
GB act.     &  6365.63 &  10817.70 &  6652.80 &  11130.79 \\
GB hist.    &  6336.83 &  11288.47 &  6269.50 &  11185.10 \\
RF act.     &   \textbf{854.42} &   \textbf{8866.62} &  \textbf{1328.32} &   \textbf{9228.00} \\
RF hist.    &  2178.95 &  10001.44 &  2499.74 &   9790.29 \\
\hline \\
\end{tabular}
\caption{Main engine installed power regression error (Mean Absolute Error) for the best configuration found for each model. Votes aggregated with mean and median.}
\label{tab:MAEpowerME}
\end{table}

To see the actual impact of the approach, the emissions are estimated with a methodology based on the STEAM model\cite{Jalkanen2009b, Jalkanen2012} using the power estimated by the best regressors found before (history and activations), the mean of the ship type and the real value. In Table~\ref{tab:emisTab} it can be observed that our approach is closer to the estimated with the real values than the best model with the original input plus history and the average by type. In fact, the proposed methodology is $152.95$ tonnes closer than the estimation using the average, detecting around $45\%$ of the otherwise undetectable emissions, also $62.15$ tonnes than the best model using the original data with history. Notice that this data-set covers 1 week of data and the emissions are evaluated over 31 ships from the test set.

\begin{figure}[h!]
	\centering
    \includegraphics[width=\linewidth]{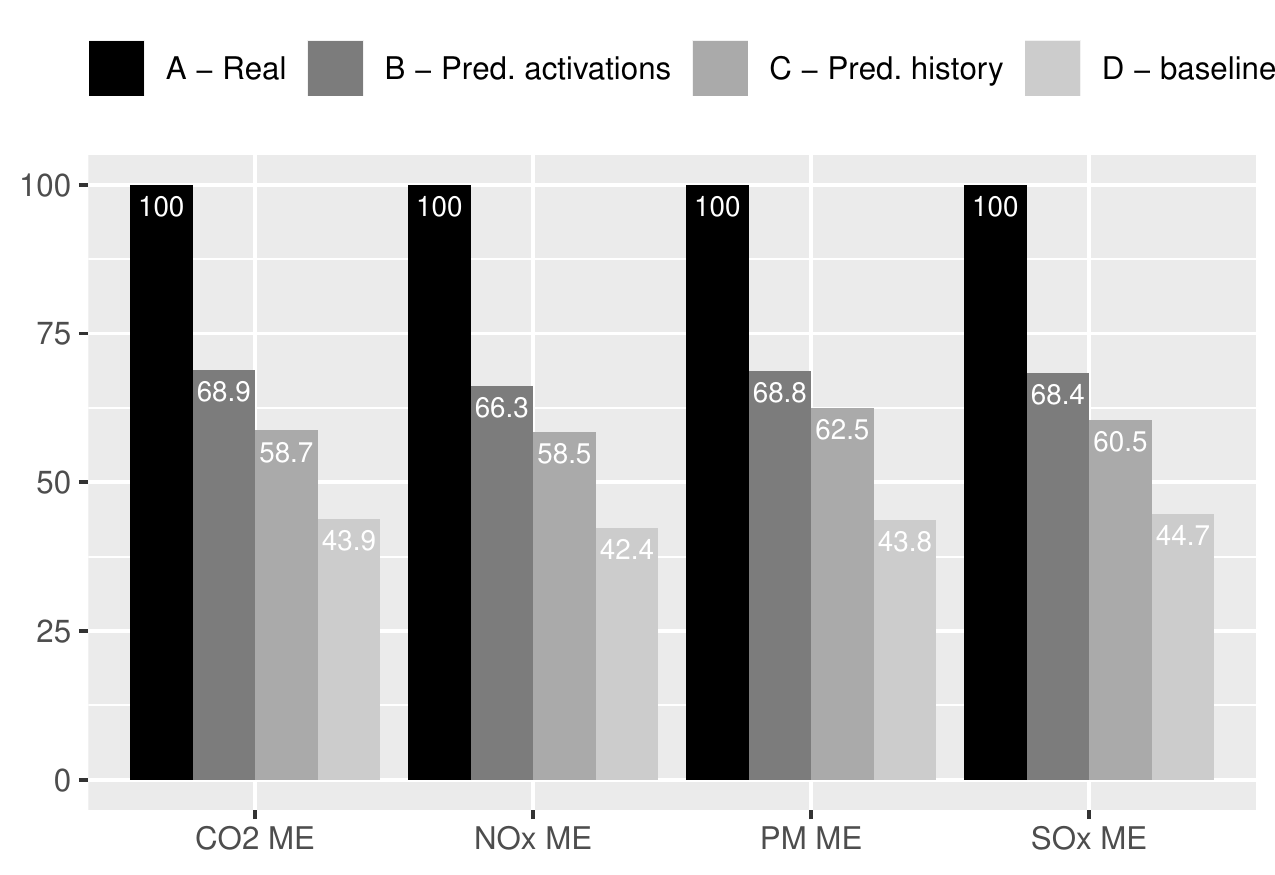}

    \caption{Percentage of pollution covered for each method and
    pollutant with the predicted main engine power. The real data marks the 100\%.}
	\label{fig:perc_emissions}
\end{figure}

In terms of the overall percentage of pollution regarding the real,  we can see that there is between an 7.9\% and 10.2\% of improvement from plain prediction method and between a 23.7\% and 25\% from the baseline in all the pollutants, as can be observed in Figure~\ref{fig:perc_emissions}.

There is still room for improvement as around a 31\% of pollution is yet to be covered, however this is not a trivial task as the variability of installed power is high.
\begin{table}[ht]
\centering
\resizebox{\linewidth}{!}{%
\begin{tabular}{rrrrr}
  \hline
  Used value & SOxME & NOxME & CO2ME & PM ME\\ 
  \hline
  Real engine values & 0.38 & 13.31 & 598.13 & 0.16\\ 
  Prediction with activations& 0.26 & 8.82 & 412.27 & 0.11 \\ 
  Prediction with history& 0.23 & 7.79 & 351.19 & 0.10\\ 
  Type average engine& 0.17 & 5.64 & 262.63 & 0.07\\ 
  \hline \\
\end{tabular}
}
\caption{Estimated pollution in tonnes for each component, using the test-set individuals with the different input values.}
\label{tab:emisTab}

\end{table}

\subsection{Navigational Status pattern mining}

The \emph{NavStatus} feature (Navigation Status) is a value manually introduced by the vessel crew. In regular cruisers or passenger boats, it is expected to be updated in a regular procedure, while most of the fishing ships do not update it and keep the same value always even though they change of operational mode. 

This attribute is essential to estimate the power usage of the auxiliary engines of the ships which are not reflected in the ship's speed, contrary to what happens with the main engine power. The ideal situation would be to use this attribute as proposed in Figure~\ref{fig:estimation_diagram}, however it is not being used directly in the emission modeling literature as it is not reliable. Instead, a 3 level operational mode surrogate variable is used. This variable is based on speed limits which define three states: moored, maneuvering and cruising. Navigational status provides more information about the usage profile of the ship, therefore it is interesting to explore this attribute and expand.

This use case proposes to focus on using the cluster labels as a surrogate for the \emph{NavStatus} indicator, to be compared to the real one and correct it when unavailable or considered more reliable. 
For this, we feed the \emph{k-means} algorithm with the \ac{CRBM} activations as previously mentioned. In this case, we selected $k = 4$ as hyper-parameter (not considering \textit{burned} samples for initial history, marked as $Cluster 1$), as lower $k$ only detected differences between movement and resting, and higher $k$ produced very similar clusters.

\begin{table*}[h!t]
\centering
\resizebox{\textwidth}{!}{%
\begin{tabular}{rrrrrrrrr}
  \hline
 & At anchor & Engaged in fishing & Moored & Not under command & Restricted maneuv. & Undefined & Under way using engine \\ 
  \hline
        1 & 0.03 & \emph{0.18} & 0.09 & 0.00 & 0.00 & 0.03 & 0.67 \\ 
        2 & 0.07 & \emph{0.17} & \textbf{0.50} & 0.00 & 0.00 & 0.04 & 0.22 \\ 
        3 & 0.02 & \emph{0.26} & 0.11 & 0.00 & 0.00 & 0.06 & \textbf{0.55} \\ 
        4 & 0.11 & \emph{0.22} & \textbf{0.47} & 0.01 & 0.00 & 0.05 & 0.14 \\ 
        5 & 0.06 & \emph{0.24} & \textbf{0.32} & 0.00 & 0.00 & 0.04 & \textbf{0.34} \\ 
   \hline
\end{tabular}
}
\vspace{1mm}
\caption{Clusters vs. NavStatus labels. Values are normalized per row. Notice that Cluster 1 refers to the \emph{delay} data not classified}
\label{tab:labelcluster}
\vspace{-5mm}
\end{table*}

Table~\ref{tab:labelcluster} shows the NavStatus vs. clustering, grouping those stopped due to anchoring and mooring, those stopped due to fishing, and those in movement. Such results allow us to validate the cluster labels:
Cluster 1, as mentioned before, is the status for the data used as initial history, not classified;
Cluster 2 refers principally to vessels mooring and in minor measure moving with their engines started, considering this maneuvering;
Cluster 3 indicates those that are moving or fishing, and we visually detected that it is assigned to those moving towards fishing positions, or it is mixed with cluster 4 in trawlers;
Cluster 4 refers to those moored or fishing, and we visually detected that such status is given to those trawling, moving much slower compared to other speeds (1/4 to 1/10 of regular moving speed);
Cluster 5 is split between moving, fishing or moored, but by visualization we observed that those labeled as 5 are actually sailing towards fishing positions or returning to port.

As fishing vessels usually set their NavStatus to ``engaged in fishing'' always, even when sailing or moored, we can determine their  ``real'' status with this classification. Also, for those without status (``undefined''), we can use the assigned cluster label as expected status, and applying approximate NavStatus labels by using the majority label for each cluster: indicating as ``moored'' if Cluster classifies it as 2, ``under way using engine'' if Cluster is 3 or 5, ``moored OR slow fishing'' if cluster is 4.

\begin{figure}[t!]
	\centering
    \includegraphics[width=.7\linewidth]{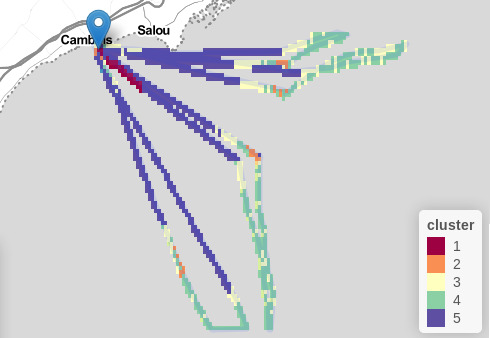}
	\vspace{-2mm}
    \caption{Example of status classification by clustering, on a trawling ship. Visualization tool at \url{http://patrons.bsc.es}.}
	\label{fig:marinaflaire}
\end{figure}

Even though the correlation of these clusters with the NavStatus is not clear, we can identify new latent behaviors. As an example, we can identify patterns for ships performing trawling, not present in other fishing ships, cargos and passenger boats. In this example, shown in Figure~\ref{fig:marinaflaire}, we can identify a first cluster (n.2) indicating the maneuvering in port and when shifting trajectories before and after trawling; two clusters (n.3 and n.4) identifying the movements during trawling, slower that regular sailing, that potentially can consume more energy thus more emissions, as they are trawling fishing nets; then a cluster (n.5) for ships speeding towards or from the fishing regions and the port.

For this identification pattern exercise, the validation has been done by expert visual recognition of ship movement traces and location according to port maps~\cite{Cepesca}, and by identifying the vessels registry indicating whether they possessed trawling equipment on board. In future research this extra status found may be used in new emission models. 

Finally, this approach shows potential to be applied in other contexts for uncovering behaviors, e.g. analyzing patterns on road traffic mining or other kind of data-set containing \ac{GPS}.

\section{Conclusions}
\label{sec:conclusions}

Computing the pollutant emissions from maritime traffic is an important issue for coastal cities air quality, as indicated by research in environmental sciences, also a major concern for world governments and global health organizations. Current state of the art techniques to model those emissions are based on processing \ac{AIS} data, from ships traces, to complement large air quality simulations currently performed in supercomputing centers. The principal problem comes when usually that data is incomplete or incorrect for a large amount of vessels.

In this paper we presented a methodology for enhancing \ac{AIS} data-sets by correcting and expanding its features, towards producing better estimations when using the latest emission models. Our proposed methods focus on using \acp{CRBM} to boost prediction and clustering algorithms, used for complete crucial missing data required for producing those estimations. 
Experiments show that ship type and navigational status may be corrected on missing data scenarios. Moreover, they show that navigational status can be expanded with new uncovered behaviors. 
Finally, Experiments have proved that our method is able to estimate emissions than those proposed by the current emission models, detecting around $45\%$ of the usually undetected emissions when the required features are not available.

Next steps will focus on the application of the produced features and methods on the emission models. Also the possibility of using the Navigational Status instead of the current approach will be further studied.

As there is still a gap between the estimated emissions with the real attributes and the corrected ones when the real are missing, future works will focus on improving the results even though the problem is hard.
Other important attributes, e.g. ship's auxiliary engine usage, will also be covered.

\section*{Acknowledgments}
We would like to thank Spanish Ports Authority (Puertos del Estado) for providing the data for this study. This project has received funding from the European Research Council (ERC) under the European Union Horizon 2020 research and innovation programme (grant  agreement  No  639595). This work is partially supported by the Ministry of Economy, Industry and Competitiveness of Spain under contracts TIN2015-65316-P, 2014SGR1051, IJCI2016-27485, and Severo Ochoa Center of Excellence SEV-2015-0493-16-5.

\bibliographystyle{elsarticle-num}
\bibliography{biblio} 

\appendix

\section{Reproducibility supplement}
In this section we give details on how to reproduce the experiments performed for this paper using the code available online \footnote{\url{https://github.com/HiEST/EmissionsMissingDataCRBM}}. Data-sets and models can be found in Patrons@BSC website\footnote{\url{http://patrons.bsc.es/datasets/}}.

\subsection{Software requirements}
In order to reproduce the results the following software is required:
\begin{itemize}
    \item R plus the following libraries: NMOF, parallel, reshape2, ggplot2, optparse, data.table, dplyr, zoo, DBI, dplyr, blob and rrbm 
    \item Python plus the following libraries: sklearn, numpy, pandas and argparse
\end{itemize}

All the R packages can be installed using R's \emph{install.packages()} function except \emph{rrbm}, which is available with install instructions on GitHub\footnote{\url{https://github.com/josepllberral/machine-learning-tools}}. The Python packages can be installed using python's pip package manager.

All the code is present in Jupyter\footnote{\url{https://jupyter.org}} notebooks but also in the corresponding .R or .py format for ease of usage. Mind that execution of Jupyter notebooks that contain R require IRKernel installation.

All the data-sets and models for each step are provided for convenience. Nevertheless, all the work can be reproduced using just the original AIS data-set (both original and cleaned) and the IHS data-set, which contains main engine power value and ship characteristics for emission estimation.

\subsection{CRBM data-sets generation}

This process consist in training a CRBM with the original data to first obtain the activations of the \ac{CRBM}, which can be used directly for the prediction experiments, and then using $k$-means to produce a clustering over those activations, as explained in this work.
In order to generate the activations and clustering data-sets we use the \ac{CRBM} implementation found in the \emph{rrbm} package. Clustering data-set requires of the calculation of the activations data-set as it is a product of applying $k$-means over it, however two different scripts are provided that perform the end-to-end case, from original data to the desired data-set, for the sake of usability. The activations data-set generator includes merging the data from IHS tables to be able to run the missing main engine experiment. Both scripts are provided in Jupyter and .R formats.

These scripts use two data-sets: the AIS data-set preprocessed (already interpolated with regular time-steps) and the IHS data-set, which provides us in this case of the installed main engine power.

\subsection{Experiments}

%

\subsubsection{Determining a valid NavStatus}
In order to reproduce this experiment the $k$-Means data-set is required. With this data, cross the two variables \emph{cluster} and \emph{navstatus} and then divide by the sum of the rows to obtain the percentages. A script that performs this procedure is provided.

The trawling ship, ship number 206, can be seen in our tool at our website. In order to see the plot present in this paper the following configuration is needed: data-set should be CRBMResults, variable cluster and ship 206.

\subsubsection{Ship Type Prediction}
In this experiment we try to find the best model to predict the ship type. In this experiment we build three different data-sets: original data (the time $t$ sample), $history$ data (the time $t$ sample plus $history$) and activations data (the result produced by the \ac{CRBM} when $history$ data is given as input). Here three models are tested: Logistic Regression, Multi-Layer Perceptron and $k$-Nearest Neighbors. The hyper-parameter search is done using \emph{sklearn}'s \emph{GridSearchCV} function for performing grid search using cross-validation. This script can be both run with Jupyter and with python directly.

For this experiment the 10 activations data-set is required. This data-set has the activations for the \ac{CRBM} trained with 10 hidden units and windowed original features so that both activations data-set and original with $history$ data-set can be evaluated and compared at the same time.

\subsubsection{Improving emission estimations on the presence of missing data}
In this experiment what we do is find the best model to predict the main engine installed power. What we do here is a loop of different models with a grid of parameters for each one, similar to the previous case. In this case we have tested Multi-Layer Perceptron and Support Vector Machines, both Radial Basis Function kernel and linear (no kernel), however these two methods were discarded because of the computational cost and accuracy trade-off. With the script we are able to build the average model (global and by type), Lasso regression, Gradient Boosting and Random Forest. All the results are saved in the specified folder. The best model, Random Forest, can be downloaded and tested with the last part of the Jupyter notebook.

For this experiment the 70 activations data-set is required. This data-set provides both CRBM and original features for convenience.

\subsubsection{Emission estimation from predicted main engine power}
We also provide an R implementation of the emission estimation methodology~\cite{Jalkanen2009b} for emission estimation. This script will use the original AIS data, provide time regularity, and estimate the emissions using the ship characteristics found in the IHS data-set and the provided file with the predicted main engine power values for Random Forest with $history$ data-set, Random Forest with activations data-set and the baseline values (average by type). This script will provide as output the emissions calculated for those three sets of values and the emissions with the real main engine values.

For this experiment the original AIS data-set, the IHS data-set and the main engine prediction data-set are required.

\end{document}